\documentclass [12pt]{article}
\usepackage{latexsym}
\usepackage{amsfonts}
\usepackage{acronym}
\newcommand{\be}{\begin{equation}}
\newcommand{\ee}{\end{equation}}
\newcommand{\bea}{\begin{eqnarray}}
\newcommand{\eea}{\end{eqnarray}}
\makeatletter
\@addtoreset{equation}{section}

 \makeatother
\begin{document}
\normalsize
\title{\Large
 Superfluid properties of BPS monopoles.}
\author
{{\bf L.~D.~Lantsman}\\
 Wissenschaftliche Gesellschaft bei
 J$\rm \ddot u$dische Gemeinde  zu Rostock,\\Augusten Strasse, 20,\\
 18055, Rostock, Germany; \\
Tel.  049-0381-799-07-24,\\
llantsman@freenet.de}
\maketitle
\begin {abstract}
This paper is  devoted to demonstrating manifest superfluid
properties of the Minkowskian Higgs model with vacuum BPS monopole
solutions at assuming the "continuous" $\sim S^2$ vacuum geometry
in that model.

It will be also argued that point hedgehog
topological defects are present in the
Minkowskian Higgs model with BPS monopoles.

It turns out, and we show this, that the enumerated phenomena are compatible with the Faddeev-Popov "heuristic" quantization of the Minkowskian Higgs model with vacuum BPS monopoles, coming to fixing the Weyl (temporal) gauge $A_0=0$ for gauge fields $A$ in the  Faddeev-Popov path integral. 

\end{abstract}
\newpage
\tableofcontents
\newpage
\section{Introduction.}
 In the series of recent works \cite{David2,David3, Pervush2,LP2,LP1} there
was worked out the model of the physical "Yang-Mills-Higgs" (YMH) vacuum in the Minkowski space involving BPS monopole solutions.

The base of the approach \cite{David2,David3, Pervush2,LP2,LP1} to constructing the Minkowskian  Higgs model was the Dirac fundamental quantization \cite{Dir} of that model, coming to the constraint-shell reduction of the theory in terms of topological  Dirac variables $\hat A^D_i$ ($i=1,2,3$) \cite{David2,David3, Pervush2} (in the YM sector of the considered Minkowskian YMH model), manifestly gauge invariant (G-invariant), relativistic covariant (S-covariant)  and transverse.

\medskip
The brief analysis of the Dirac fundamental quantization method \cite{Dir} was performed recently in Ref. \cite{fund}.

The historical retrospective development the Minkowskian Higgs model quantized by Dirac \cite{Dir}  was given in \cite{fund}, and principal results  got in \cite{David2,David3, Pervush2,LP2,LP1} about the Dirac fundamental quantization of the Minkowskian  Higgs model involving vacuum BPS monopoles were looked into.

Also the principal distinctions between the Dirac fundamental quantization method \cite{Dir} and the  Faddeev-Popov (FP) "heuristic" gauge fixing method \cite{FP1} were pointed out in \cite{fund}.

Repeating the arguments \cite{Arsen}, there was demonstrated that these come, basically, to violating the gauge equivalence (independence) theorem \cite{Taylor, Slavnov1} in the cases of collective vacuum excitations and bound states.

\medskip
On the other hand, as it was noted already in \cite{fund} (towards the end of the discussion therein), the investigations about the Minkowskian  Higgs model quantized by  Dirac and  involving vacuum BPS monopoles are far today for their finishing,
and
there lot of job is in prospect in order to specify details the  model  constructed in \cite{David2,David3, Pervush2,LP2,LP1} and to made new observations.

\medskip
Just specifying  details of the Dirac fundamental quantization \cite{Dir} for the Minkowskian  Higgs model with BPS monopoles will be devoted the series of papers we now begin.

In the present paper we shall occupied ourselves with the more profound analysis of the Minkowskian  Higgs model with BPS monopoles, digressing for some time
from assuming about the Dirac fundamental quantization \cite{Dir} of that model.

We continue to study the "classical" Minkowskian  Higgs model with BPS monopoles, the job beginning in Refs. \cite{LP2,LP1}.

Herewith speaking about the "classical" Minkowskian  Higgs model with BPS monopoles, we mean results got without solving the Gauss law constraint, issuing only from the action functional of the Minkowskian  Higgs model before fixing any gauge.

A good analysis of the "classical" Minkowskian  Higgs model with BPS monopoles was performed in Refs. \cite{Al.S.,BPS, Gold} (this analysis was reproduced partially in \cite{LP2,LP1}).

The important premise for the "classical" Minkowskian  Higgs model \cite{Al.S.,BPS, Gold} with BPS monopoles is assuming about the "continuous"
$$   R\equiv SU(2)/U(1)\simeq S^2$$
vacuum geometry.

Repeating the reasoning \cite{Al.S.}, we show in {\it Section 1} that such "continuous" vacuum geometry implies point (hedgehog) topological defects presenting in the "classical" Minkowskian  Higgs model \cite{Al.S.,BPS, Gold} with vacuum BPS monopole solutions.

\medskip
The next important topic of the present study will be demonstrating manifest superfluid properties the "classical" Minkowskian  Higgs model with BPS monopoles. This will be done in {\it Section 2}.

These properties, unique for the "classical" Minkowskian  Higgs model \cite{Al.S.,BPS, Gold} with vacuum BPS monopole solutions, are induced by the Bogomol'nyi equation \cite{Al.S.}
$$  {\bf B} =\pm D \Phi,      $$
giving the relation between the vacuum "magnetic" field $\bf B$ and the Higgs isomultiplet $\Phi$.

We argue  that the Bogomol'nyi equation is just the {\it potentiality condition} for the vacuum of the "classical" Minkowskian  Higgs model with BPS monopoles.

The transparent parallel   of the Higgs model \cite{Al.S.,BPS, Gold} and the liquid helium II (at rest) theory \cite{N.N.} will prove to be helpful for us in this argumentation.

The enumerated properties of the "classical" Minkowskian  Higgs model with BPS monopoles (we mean point hedgehog topological defects and manifest superfluid properties inherent in that model) are compatible with the FP "heuristic" quantization \cite{FP1} of that model.

We demonstrate (and this will be the one of most important topics of the present study) that the FP "heuristic" quantization \cite{FP1} of the Minkowskian  Higgs model with vacuum BPS monopole solutions comes to fixing the gauge $A_0=0$ for YM fields in the appropriate FP path integral.

\medskip  Additionally, we propose to our readers an important Appendix where we give the mathematical theory of magnetic charge $\bf m$ (repeating the arguments stated in the monograph \cite{Al.S.}). This theory also closely related to the natural isomorphism 

$$\pi_2 R= \pi_1 H $$ for the vacuum manifold $R$ and residual (gauge) symmetry group $H$ in the models of such kind we discuss in the present study. Just such  isomorphism generates pint topological defects (magnetic monopoles) inside the vacuum manifold $R$.
\section{Point hedgehog topological defects always accompany  "continuous" vacuum geometry.}
Let us denote as $G$ the initial symmetry group in a gauge (Minkowkian) model.

If this model implicates the spontaneous breakdown  of the initial gauge symmetry group $G$ (as a rule, this is associated with Higgs modes), we shall denote as $H$ the appropriate residual gauge symmetry group.

For instance, in the Minkowskian  YMH model
$$ G\equiv SU(2), \quad  H\equiv  U(1),     $$
respectively.

As it was shown in the monograph \cite{Ryder} (and these arguments were repeated in \cite{LP2,LP1}), in this case the initial gauge symmetry group $G$ may be represented as
\be
\label{sxema1}
G=H \oplus G/H.\ee
Herewith the second item in (\ref{sxema1}), $R\equiv G/H$, is, from the geometrical viewpoint, is a space proving to be invariant under  gauge transformations $H$.

Additionally, it may be supposed that
$$ H R_i= R_i     $$
for all the points $R_i$ of this space:
in other words, that that $H$  is the \it stationary subgroup \rm of the point $R_i\in R$.  \par
In particular, in the Minkowskian  YMH model
\be \label{S2}
R= SU(2)/U(1) \simeq S^2.\ee
The space $R$ is called the {\it vacuum manifold}.

This term is justified by two considerations.

Firstly, vacuum manifolds in gauge models may be defined merely as those invariant with respect to appropriate (residual) gauge groups in these models.

Good explaining this fact was given in the monograph \cite {Ryder}, in \S 8.1 \footnote{In this case the following general definition \cite {Al.S.} of quantum fluctuations over a vacuum manifold $R$ may be given.

Such quantum fluctuations belong to a set $\cal R$ treated as an (infinitesimal) neighbourhood of the appropriate vacuum manifold $R$.}.

The above remark prompts the possibility to give an alternative interpretation of vacuum manifolds. Such interpretation was given in Ref. \cite {Al.S.}, in \S$\Phi$1.

It is quite correct to interpret  the initial symmetry group $G$ as that does not change the energy functional (Hamiltonian) of the considered (e.g. non-Abelian) theory, while the residual symmetry group $H\subset G$ as that consisting of transformations  that keep invariant a fixed equilibrium state. \par
All these states (at a fixed temperature $T$ \footnote{If the initial symmetry $G$ in the considered gauge theory is violated up to its subgroup $H$, $T<T_c$, with $ T_c $ being the Curie point in which the initial symmetry $G$ is violated and the second-order phase transition occurs. \par

It will be useful to adduce here some evidences in favour the fact  that second-order phase transition occur indeed in Minkowskian Higgs models with YM fields (YMH models). \par
To do this, let us write down (following \cite {LP1}) explicitly the appropriate action functional. This has the typical look
$$ S=-\frac {1}{4 g^2} \int d^4x F_{\mu \nu}^b F_b^{\mu \nu }+ \frac {1}{2} \int d^4x (D_\mu\Phi,D^\mu\Phi
) -\frac {\lambda}{4} \int d^4x \left[(\Phi^b)^2- \frac{m^2}{\lambda}\right]^2,
    $$
with 
$$D_\mu\Phi=\partial^\mu\Phi+g[A^{\mu },\Phi]$$ 
being the covariant derivative an $g$ being the YM coupling constant.\par
The action functional (\ref{YM L}) results the equations of motion \cite{Cheng}
$$ (D_\nu F^{\mu \nu})_a=-g \epsilon_{abc}\phi^b  (D_\mu\phi)^c,      $$
$$ (D^\mu D_\mu \Phi)_a =-\lambda \Phi_a ({\vec \Phi }\cdot{\vec \Phi }-a^2) ;\quad a^2=m^2/\lambda.
 $$
Just the second of these equations of motion implies the second-order phase transition occurring  in Minkowskian YMH models. \par 
Issuing from this equation, one can demonstrate, repeating the arguments \cite {Linde} (see \S3.1 in this monograph) that the {\it false} vacuum $\Phi =0$ may be linked in a continuous (although not smooth) wise with the {\it true} vacua $\Phi =\pm a$.\par
More exactly, at  finite temperatures $T\neq 0$, if a shift of the Higgs field: $\Phi \to\Phi(T) +\delta\Phi$, is performed, it turns out that the equation of motion for $\delta\Phi $ can be recast to the look \cite {Linde}
$$ (D^\mu D_\mu \delta\Phi)_a -[-m^2+(\lambda/4)T^2] \delta\Phi_a=0,  $$
got in an infinitesimal neighbourhood of $\Phi(T)=0 $ at utilizing the relations 
$$ <\Phi ^2>\sim T^2/12$$
for the Gibbs average of the bosonic field $\Phi $ squared, and
$$   \Phi(T) =\sqrt{a^2-T^2/4}.  $$
Such  field is the one of solution (together with $\Phi(T)=0$) of Eq. \cite {Linde}
$$  \Phi(T)~ [\lambda\Phi^2(T)-m^2+(\lambda/4)T^2]=0.   $$
In turn, this is the look of the general equation
$$  (D^\mu D_\mu \Phi)_a(T)- [\lambda\Phi^2(T)-m^2+(\lambda/4)T^2] \Phi(T) =0,      $$
in which a constant value ($D^\mu D_\mu \Phi=0$) is substituted. \par
Generally speaking, to get the equations of motion in Minkowskian YMH models at finite temperatures, the Gibbs average of these equations would be taken \cite {Linde}.\par
 Above Eqs. are the particular case where this method is applied. \par\medskip 
We can assert now that at finite temperatures $T\neq 0$ the Higgs field $\Phi $ acquires the effective mass 
$$  m'=-m^2+  (\lambda/4)T^2.   $$
in the point $\Phi(T) =0$;  then \cite {Linde}
$$ (D^\mu D_\mu \delta\Phi)_a -m'^2\delta\Phi_a=0.$$ 

The value $m'^2$ becomes negative at
$$ T<T_c; \quad T_c=2m/\sqrt\lambda \equiv 2a.$$
Vice verse, it becomes positive at $T>T_c$.\par 
In a point $\Phi(T) \neq0$ (in an infinitesimal neighbourhood of $\Phi(T) =0$),  $m'^2(\Phi,T) >0$ \linebreak for $\Phi(T) =\sqrt{a^2-T^2/4}$ \cite {Linde}:
$$ m'^2(\Phi,T) =3\lambda \Phi^2(T) -m^2+  (\lambda/4)T^2 =2\lambda \Phi^2(T). $$ 
To derive this Eq., it is necessary to take account of $<\Phi> =0$ \cite {Linde}. Then $\delta(\Phi^3) =3\Phi^2\delta \Phi $ would be substituted  for deriving $m'^2(\Phi,T)$. \par 
Thus the solution $\Phi(T) =\sqrt{a^2-T^2/4}$ to the YMH equations of motion at finite temperatures is steady at $T<T_c $ and vanishes (more precisely, becomes complex, losing thus its physical sense) at $T>T_c$, in the moment when the solution 
$\Phi =0$ (the false vacuum solution) becomes steady. \par 
This means that a phase transition occurs at the temperature (Curie point) $T=T_c$ implicating restoring the (initial) $SU(2)$ gauge symmetry. \par 
The results have been got may be illustrated graphically. \par
It turns out \cite {Linde} that the lines $\Phi(T) =0$ (coinciding with the abscises axis $T$) and $\Phi(T) =\sqrt{a^2-T^2/4}$ are linked in a continuous (although not smooth) wise. \par
This just corresponds to the second-order phase transition occurring in the Minkowskian YMH model. \par
\medskip
Note that {\it first-order}  phase transitions take place in some (Minkowskian) Higgs models: in particular in the  Minkowskian Abelian Higgs model and in GUT. \par
This was demonstrated in \cite {Linde}. The principal argument that was suggested in \cite {Linde} in favour of first-order  phase transitions occurring in the mentioned models are gaps  that appropriate plots $\Phi(T)$ suffer (more precisely, it is impossible to link in a continuous wise the false vacuum $\Phi(T)=0$ and other, {\it stable}, vacua in the quested models). \par
For example, in the  Minkowskian Abelian Higgs model with the Lorentz gauge $\partial_\mu A^\mu=0$ fixed the first-order  phase transitions occurs when $\lambda\leq e$ (with $e$ being the elementary charge).
In this case it can be shown \cite {Linde} that three solutions to appropriate equations of motion exist, including the false vacuum $\Phi(T)=0$, in the temperatures interval $T_{c_1}<T< T_{c_2}$.\par

As a result, the false vacuum $\Phi(T)=0$ becomes a metastable state at $T> T_{c_1}$, while there exists also the instable   solution $\Phi_2$ corresponding to the local maximum of the  appropriate potential $V(\Phi,T)$ and $\Phi_1$ corresponding to the "true" vacuum, i.e. to the (global) minimum of $V(\Phi,T)$.  
Herewith the phase transition from the $\Phi_1$ to the $\Phi(T)=0$ state, accompanied by restoring the $U(1)$ gauge symmetry, begins at a temperature $T_c$ at which \cite {Linde}
$$ V(\Phi_1(T_c),T_c)\sim V(0,T_c).       $$
Graphically \cite {Linde}, there is however a gap between the both potentials at $T=T_c$, corresponding to the gap $\Delta F(T)=\Delta V(\Phi,T)$ (the {\it latent heat}) in the free energy $F(T)$. It is just the sign of the  first-order  phase transition occurring in the  Minkowskian Abelian Higgs model. \par
On the other hand, the metastability of the false vacuum $\Phi(T)=0$ implies the {\it supercooling } phenomenon: the system of fields remains in this state, coexisting simultaneously with the "true" vacuum $\Phi_1$ even when $T<T_c$, coexisting herewith simultaneously with the "true" vacuum $\Phi_1$. 
Vice verse, when $T>T_c$, the latent  heat is liberated as  $\Delta F(T)$. This phenomenon is referred to as {\it reheating} \cite {Linde,Coles}.
 })  form the so-called \it degeneration space \rm (vacuum manifold), that we just denote as $R$ in the present study, following  \cite {Al.S.}. \par
The natural claim to this space in gauge theories is herewith  \it to be topological\rm. \par

\medskip
The structure of a degeneration space may be investigated with the aid of the
Landau  theory  of second-order phase transitions (see e.g. \S 142 in \cite{Landau5}). \par
An equilibrium state is determined by the condition for the
free energy of the given system  to be minimal. \par
In the
Landau  theory  of second-order phase transitions (the pattern of which is the  Minkowskian Higgs model) one supposes that an equilibrium state may be found at minimizing  the free energy of the given system  by the set of states   specified by a finite number of parameters (called \it  order parameters\rm), but not by the set of all the states. \par

\medskip
Return to the case of the  Minkowskian YMH  model, when
\be
R= \label{contin} SU(2)/U(1)\simeq S^2.\ee
In this case, obviously,
\be
\label{point1}
\pi_2 (R)= \pi_2 S^2 = \pi_ 1(H)= \pi_ 1 U(1)=\bf Z. \ee
Generally speaking, repeating the arguments  \cite{Al.S.} (\S$\Phi$1), there may be shown that the isomorphism (\ref{point1}) predestines the existence of \it point topological defects \rm in a gauge theory with the spontaneous breakdown of the initial gauge symmetry $G$ down to its subgroup $H$, involving the vacuum manifold $R=G/H$.
\par
The ground cause of these topological defects is violating  the thermodynamic  equilibrium over an (infinitesimal) neighbourhood $U$ of a point in the coordinate (e.g. in the \linebreak Minkowski) space. \par
Such neighbourhood is topologically equivalent to the two-sphere $S^2$ \footnote{Indeed, a two-sphere $S^2$ may be always picked out inside such $U$: $ S^2\in U$. In this case, following \cite{Al.S.}, let us denote simultaneously as $f$ two maps.
\par
These are, firstly, the map $f:U\to R$ and, secondly, the map $f:S^2\to R$.
Latter one is treated as the restriction of the map $f:U\to R$ onto the sphere $ S^2$.\par
If the map $f:S^2\to R$ is not homotopical to zero ($\pi_2 (R)\neq 0$), this map cannot be continue  onto the map $D^3\to R$, with $D^3$ being the ball restricted by the sphere $ S^2$ (the complete proof of the latter fact was given in Ref. \cite{Al.S.}, in \S T1).
\par
Just the said means that  there is a point topological defect inside  the sphere $ S^2$\rm. }.
\par
\medskip
Already mentioned violating  the thermodynamic  equilibrium in a neighbourhood \linebreak $U\simeq S^2$ (is the case of  point topological defects inside the vacuum manifold $R$) is always associated with gaps (singularities) which order parameters inherent in appropriate gauge models suffer over such infinitesimal neighbourhoods $U$ that may be contracted into points. \par

As the typical order parameter in Minkowskian Higgs models involving vacuum monopoles (the model \cite{Al.S.,BPS, Gold} with BPS monopoles is the one of such models), the  vacuum "magnetic" field $\bf B$ appears.

The direct computations have been performed in Ref. \cite{BPS} for the Minkowskian Higgs model with vacuum BPS monopoles results the $O(r^{-2})$ behaviour of the vacuum "magnetic" field $\bf B$ at the origin of coordinates.

The same $O(r^{-2})$ behaviour of the vacuum "magnetic" field $\bf B$ at the origin of coordinates can be observed in two another very important Minkowskian models with  monopoles.

There are the Wu-Yang  model \cite{Wu} (analysed in detail in Refs.  \cite{Pervush2,LP2,LP1}) and the 't Hooft-Polyakov model \cite{H-mon, Polyakov} \footnote{Indeed, as it was explained in Refs.  \cite{Pervush2,LP2,LP1}, Wu-Yang  monopoles $\Phi_i $ arisen in the Minkowskian model \cite{Wu} are solutions to the classical equation of motion
$$ D^{ab}_k(\Phi_i)F^{bk}_a(\Phi_i)=0       $$
of the  "pure" (Minkowskian) YM theory (without Higgs fields).

Unlike Wu-Yang  monopoles \cite{Wu},'t Hooft-Polyakov monopoles \cite{H-mon, Polyakov} are specific vacuum solutions of the Minkowskian Higgs model in its YM and Higgs sectors (see \S.10.4 in \cite{Ryder}).

And furthermore, 't Hooft-Polyakov monopoles \cite{H-mon, Polyakov} are associated with the Georgi-Glashow  theory involving the initial $SO(3)\simeq SU(2)$ gauge symmetry violated then down to its $U(1)$ subgroup (this model was experimentally ruled out after the discovery of neutral-current phenomena.).}.

The said about the $O(r^{-2})$ behaviour at the origin of coordinates of the vacuum "magnetic" field $\bf B$, treated as the order parameter in the enumerated Minkowskian models with vacuum monopole solutions, just testifies in favour of point topological defects inside appropriate vacuum manifolds (topologically equivalent to the two-sphere $S^2$).

The kind of point topological defects located at the origin of coordinates is referred to as {\it point hedgehog topological defects} \cite{Al.S.}.

This terminology is connected historically with {\it Polyakov hedgehogs}  \cite{Ryder, Polyakov}, Higgs solutions in the 't Hooft-Polyakov model \cite{H-mon, Polyakov}:
\be
\label{hedg}
\phi^a \sim \frac {x^a}{r}f(r,a),
\ee
involving a continuous function $ f(r,a)$ (with $a$ being the radius of the two-sphere $R\simeq S^2$, (\ref{S2})) \footnote{The look (\ref {hedg}) for Polyakov hedgehogs was cited in the monograph \cite{Linde} (p. p.
114- 116).}.

\bigskip
The vacuum monopole solutions in Minkowskian (Higgs) models us cited in the present study may be got issuing from {\it constrained} action functionals, i.e. before solving the YM Gauss law constraint \cite{fund}
\be \label {Gau} \partial W/\partial A_0=0\ee
with fixing a gauge.

More concretely, in the 't Hooft-Polyakov model \cite{H-mon, Polyakov}, Higgs and YM monopole solutions: there are Polyakov hedgehogs (\ref{hedg}) in the Higgs sector of that model and \cite{Ryder}
\be
\label{Polyakov p}
A_i^a= -\epsilon _{iab} \frac{r^b}{gr^2}, \ee
in the YM sector,
are derived as solutions to the equations of motions \cite{Ryder}
\be \label{eom1}
(D_\nu F^{\mu \nu})_a=-g \epsilon_{abc}\phi^b  (D_\mu\phi)^c,
\ee
\be \label{eom2}
(D^\mu D_\nu \phi)_a =-\lambda \phi_a ({\vec \phi }\cdot{\vec \phi }-a^2).
\ee
(with $\lambda$ being the Higgs selfinteraction constant); $a=m/\sqrt\lambda$ (with $m$ being the mass of the  Higgs field) is the radius of the  two-sphere $R\simeq S^2$, (\ref{contin}).

These, in turn, follow immediately from the standard action functional \cite{LP2,LP1}
  \be \label{YM L}
S=-\frac {1}{4 g^2} \int d^4x F_{\mu \nu}^b F_b^{\mu \nu }+ \frac {1}{2} \int d^4x (D_\mu\phi,D^\mu\phi
) -\frac {\lambda}{4} \int d^4x \left[(\phi^b)^2- \frac{m^2}{\lambda}\right]^2
\ee
of the  Minkowskian Higgs model.

\medskip
In the Minkowskian Wu-Yang model \cite{Wu} the equation of motion \cite{LP2,LP1}
\be \label{eq} D^{ab}_k(\Phi_i)F^{bk}_a(\Phi_i)=0 \Longrightarrow \frac {d^2f}{d r^2}+\frac {f (f^2-1)}{r^2}=0
\ee
 of this purely YM model result Wu-Yang "ansatzes" $f=\pm 1$ (at $r\neq 0$) corresponding to Wu-Yang monopoles $\Phi_i $ with topological charges $n=\pm 1$, respectively (hedgehog and antihedgehog in the terminology \cite{Polyakov}).

\medskip
Unlike the ’t Hooft-Polyakov model, in the "classical" Minkowskian Higgs model \cite{Al.S.,BPS, Gold} with vacuum BPS monopole solutions,  the latter are, indeed, solutions to the {\it Bogomol'nyi equation}  \cite{LP2,LP1,Arsen,Al.S.,BPS, Gold}
\be
\label{Bog}
{\bf B} =\pm D \Phi,
\ee
derived (see \S$\Phi$11 in \cite{Al.S.}) at evaluating the  {\it Bogomol'nyi bound}
\be
\label{Emin}
E_{\rm min}= 4\pi |{\bf m }|\frac {a}{g},~~~~~~~~~~~~\,\, ~~~~~a=\frac{m}{\sqrt{\lambda}};
\ee
with $\bf m$ denoting the magnetic charge, of the YMH field configuration energy \footnote{The ground, why we write $ |{\bf m }|$ here will be explained below.}.

The essential point  deriving the Bogomol'nyi equation (\ref{Bog}) is \cite{LP2,LP1,Al.S.} going over to the {\it Bogomol'nyi-Prasad-Sommerfeld} (BPS) limit
\be
\label{lim}
\lambda\to 0,~~~~~~m\to 0:~~~~~~~~~~ ~~~~~\frac{1}{\epsilon}\equiv\frac{gm}{\sqrt{\lambda}}\not =0.
\ee
\medskip
All these Minkowskian models with monopoles may be described with the aid of the FP path integrals formalism \cite{FP1}.

Herewith it is enough \cite{Al.S.} to fix the temporal (Weyl) gauge $A_0=0$ for YM fields via the Dirac delta-function $\delta(A_0)$ in the appropriate FP path integrals.

This  just corresponds to $F_{0i}^a=0$ for "electric" fields in the enumerated Minkowskian models with stationary monopole solutions.

\bigskip  It will be now very constructively to derive in detail the  Bogomol'nyi equation (\ref{Bog}).  Note firsr that it can be recasted in the tensor shape \cite{Al.S.}:
\be 
\label{Bog1}
 \frac {1}{2g}\epsilon ^{ijk}F_{jk}^a =\nabla ^i\Phi^a. 
\ee
Then, following  to the  't Hooft-Polyakov model \cite {H-mon, Polyakov}, let us introduce the "electromagnetic tension" as the scalar product
\be
\label{tens} 
F_{\mu\nu}= <F_{\mu\nu}^a, \frac {\Phi_a}{a} >. 
\ee
Note incidentally (and this will be later on important for us) that $\Phi/{a}$
my be treated \cite{Al.S.} as the normalized generator of the $U(1)$ group, residual gauge symmetry in the 
Minkowskian YMH theory involving (vacuum) BPS monopole solutions. Note also that  multiplying  $\Phi/{a}$
by a continuous function does not change the matter:  such a value also my be treated as the generator of the $U(1)$ group. \par 

The magnetic tension corresponding to the tensor (\ref{tens}) is 
\be
 \label{H} B^a= \frac {1}{2} \epsilon ^{ajk} <F_{jk}^b,\Phi_b > a^{-1}. \ee  
One can  define a magnetic charge $\bf m$ (in a Minkowskian YMH theory) as the  flux of the magnetic
tension $\bf B$ through the given space-like surface \rm
(multiplied by $(4\pi)^{-1}$): 
\be
\label{mul} 
{\bf m} =\frac {1}{4\pi} \int d{\bf S}~ {\bf B} = \frac {1}{8\pi}\int d^3 x ~\partial _i \{\epsilon ^{ijk} <F_{jk}^b,\Phi_b >a^{-1}\}.
\ee 
Note also that
$$ \epsilon ^{ijk} \partial _i<F_{jk}^b,\Phi_b > = \epsilon ^{ijk} \nabla _i <F_{jk}^b,\Phi_b > =$$ 
$$ \epsilon ^{ijk} (<\nabla _i F_{jk}^b,\Phi_b >+ <F_{jk}^b,\nabla _i\Phi_b>)= \epsilon ^{ijk}<F_{jk}^b,\nabla _i\Phi_b> $$ 
(we have utilized here the fact that the usual derivative $\partial _i$
coincides with the covariant derivative $\nabla _i$ for the gauge invariant (scalar) value $<F_{jk}^b,\Phi_b >$; we  also have taken  account of the Bianchi identity $\epsilon ^{ijk}\nabla _i F_{jk}^b =0$). \par 
Therefore 
\be \label{m1} {\bf m }=\frac {1}{8\pi}\int d^3 x ~
\epsilon ^{ijk}<F_{jk}^b,\nabla _i\Phi_b>a^{-1}. \ee
Then we consider the inequality
 \be 
\label{in}
\int  dx ~<c,b> ~\leq ~ \frac {1}{2} \int dx~(<c,c>+<b,b>),
\ee 
following from the relation 
$$\int <c-b,c-b>dx  \geq 0;$$  the equality
is achieved only in the case $c(x)=b(x)$ (here $c(x)$ and $b(x)$ take theirs values in ${\bf R}^n$).\par 
Applying the inequality (\ref{in}) to the tensors $\frac {1}{2g}\epsilon ^{ijk}F_{jk}^b$ and $\nabla _i\Phi_b$, we get
 \be 
\label{est}
 \int d^3x~\frac{\epsilon^{ijk}}{2g}<F_{jk}^b,\nabla _i\Phi_b> ~ \leq ~ \frac {1}{2}\int d^3x\{ \frac {1}{4g^2}<F_{jk}^b,F_{jk}^b> + <\nabla _i\Phi_b,\nabla _i\Phi_b>\}. \ee Herewith the {\it equality}  is achieved namely when the Bogomoln’nyi equation in the shape (\ref{Bog1}). 

To prove this, let us consider the square $$< \frac{1}{2g}\epsilon^{ijk}F_{jk} - \nabla^i\Phi, \frac{1}{2g}\epsilon^{il m}F_{l m} - \nabla_i\Phi >~ \geq 0$$.  Let us expand this square:  $$\frac{1}{4g^2}\langle F_{jk},F_{jk}\rangle 
+
\langle \nabla_i\Phi,\nabla_i\Phi\rangle -
\frac{1}{g}\epsilon^{ijk}\langle F_{jk},\nabla_i\Phi\rangle \geq 0. $$ Move the last term to the right:

\be \label{rel}  \frac{1}{g}\int d^3x,\epsilon^{ijk}\langle F_{jk},\nabla_i\Phi\rangle \leq \int d^3x[\frac{1}{4g^2}\langle F_{jk},F_{jk}\rangle + \langle \nabla_i\Phi,\nabla_i\Phi\rangle].\ee Now we multiply the both sides of Eq. (\ref{rel}) onto $1/2$ and get
$$\int d^3x, \frac{\epsilon^{ijk}}{2g}\langle F_{jk},\nabla_i\Phi\rangle \leq \frac{1}{2}\int d^3x[\frac{1}{4g^2}\langle F_{jk},F_{jk}\rangle + \langle \nabla_i\Phi,\nabla_i\Phi\rangle].  $$  Herewith the equality is achieved only if $$ \frac{1}{2g}\epsilon^{ijk}F_{jk} = \nabla^i\Phi.$$ And this is just the desire Bogomol’nyi equation!

\medskip

 The integral on  the right-hand side of (\ref{est}) differs from the energy $E$ of the YMH configuration $(\Phi^a, A_\mu^a)$ only in the absence
of the potential term: 
\be
\label{self} E_1= \frac {1}{4}\lambda \int d^3x~ [\Phi^2 - a^2]^2. \ee 
Since the left-hand side of (\ref{est}) differs only in the factor from the magnetic charge $\bf m$, we get  then 
\be \label{est1}
{\bf m} \leq \frac {g}{4\pi a} (E- E_1).
\ee 
In other words, \be 
\label{h.neq} 
E\geq 4\pi{\bf m} \frac {a}{g} +E_1, 
\ee 
and the equality  is achieved only in the case (\ref{Bog1}). \par 
So long as $E_1\geq 0$ and the magnetic charge $\bf m$ takes only the integers in the normalization us adopted, the energy $E$ of the YMH  configuration $(\Phi^a, A_\mu^a)$ allows the estimation \be 
\label{est3} E\geq 4\pi|{\bf m}|\frac {a}{g} 
\ee
in a general case. Here an important caution is appropriate. In the case of a negative magnetic charge, it should be taken $|{\bf m}|$ since  for negative ${\bf m}$ we encounter a negative lower bound, which is meaningless physically. The zero magnetic charge is permissible.\par 
In the BPS limit \cite{Al.S.,BPS}, $\lambda \to 0$,  $m\to 0$, if the other parameters: $a$ and $g$, remain invariable, this estimation becomes exact:
\be 
\label{energy} 
E= \frac {1}{4g^2} \int d^3x~ <F_{jk}^b,F_{jk}^b>  +\frac {1}{2}\int d^3x~ <\nabla _i\Phi_b,\nabla _i\Phi^b>. \ee 
If the fields $(\Phi^a, A_\mu^a)$ satisfy the  Bogomol'nyi equation (\ref{Bog}), (\ref{Bog1}),  the functional  (\ref{energy}) 
achieves
 its minimum (the Bogomol'nyi bound) (\ref{Emin}).\par 

Note that in this case, as ${\bf m}=0$, one  comes to the vacuum BPS monopole solutions \cite{LP2,LP1,Al.S.} in the Higgs and YM sectors. 
\par

\bigskip In the here discussed model  \cite{Al.S.,BPS,Gold}  involving vacuum BPS monopole modes, we consider the topologically degenerated vacuum BPS-monopole configuration  $(\Phi^a ({\bf x}), A_\mu^a({\bf x}))$  in a fixed time instant $t_0$ \cite{Al.S.}. However, the magnetic charge $\bf m$ does not depend on a choice of this time instant (this directly follows from the invariability of the magnetic charge (\ref {mul}) at continuous deformations of the YM field $ A_\mu$). Thus we can speak about the magnetic charge
of the  vacuum BPS-monopole configuration 
$(\Phi^a ({\bf x}), A_\mu^a({\bf x}))$ contained inside  a  surface $\Gamma$ (in our case it is the space-like surface $t= t_0$ in the Minkowski space-time\rm) being in a one-to-one
correspondence with the  appropriate vacuum manifold $M_0$.  

\medskip The shape of such vacuum manifold  depends essentially on the model we study. So, for instance, in the ’t
Hooft-Polyakov model \cite{H-mon,Polyakov}, it is merely the    minimum manifold of the potential $V$. We have encountered already this potential above. Let us now write it out once again. It is 
\be \label{potential} V=\frac {\lambda}{4} \int d^4x \left[(\Phi^b)^2- \frac{m^2}{\lambda}\right]^2.
\ee  In this framework, the   vacuum manifold $M_0$ takes the shape
\be 
\label{min}
 M_0=\{\vert \vec\Phi\vert= a;\quad a^2=m^2/ \lambda\} 
\ee 
as ${\bf r} \to \infty$. Thus $M_0$ consists of the points of the sphere $S^2$ in the three-dimensional $SU(2)$ group space. \par 

\medskip In the case of the Dirac fundamental quantization of the Minkowskian YMH model involving Higgs and YM BPS vacuum modes, all things become more  complex. This is associated with the need to take into account the nontrivial topological dynamics inherent in this model (we shall say a few words about this below). This implies the so-called {\it discrete vacuum geometry} providing point (of the hedgehog kind) and thread topological defects inside the vacuum manifold. See the review \cite{disc} for details.

\medskip We can write down now \be \label{potok} {\bf m}_ \Gamma = (4 \pi)^{-1} \int \limits _ \Gamma  {\bf B}~ d{\bf S}\ee
in a general case. \par 
It is  easy to check that ${\bf m}_ \Gamma $ is invariant  with respect to a continuous deformation of the YM field $ A_\mu$,
if the values of the Higgs field $\Phi ({\bf x})$ belong to the vacuum manifold at this deformation.  \par
Due to (\ref {H}), (\ref {mul}), the magnetic charge
of the given  vacuum BPS-monopole configuration 
$(\Phi^a ({\bf x}), A_\mu^a({\bf x}))$ is completely
 specified  by the Higgs field $\Phi ({\bf x})$. \par
A configuration $(\Phi^a ({\bf x}), A_\mu^a({\bf x}))$ has the same magnetic charge that another configuration $(\Phi^a ({\bf x}), \tilde A_\mu^a({\bf x}))$ if one may link 
them by a continuous deformation  
$$(\Phi^a ({\bf x}), ~~t\tilde A_\mu^a({\bf x})+(1-t) A_\mu^a({\bf x})), \quad 0\le t\le 1.$$  
Thus it is easy to see that the magnetic charge $\bf m$ of the  vacuum BPS-monopole configuration 
$(\Phi^a ({\bf x}), A_\mu^a({\bf x}))$ is a function of the topological charge $\bf n$.

Moreover (see \S$\Phi$7 in \cite{Al.S.}),   it can be shown that  a magnetic charge $\bf m$ is a linear   function of the topological charge:
\be  
\label{mt1} 
{\bf m} (\Phi,A)= C~ \zeta (\Phi,A), \quad \zeta (\Phi,A)\in {\bf Z}, \ee  This follows from the  fact that  the magnetic charge is additive; the topological charge possesses the same property. \par 
Let us now prove that $C=\nu/4\pi$, where $\nu$ can be found  from the conditions 
\be  
\label{norm.usl.} 
\exp (\nu h)=1; \quad \exp (\lambda h)\neq 1 
\ee 
($h\equiv h(\Phi) \equiv \Phi /{a}$) as $0\leq \lambda< \nu$.  One also  can speak that $\nu$ is the minimal positive number for which 
$\exp (v h)=1$. 
From the geometrical point of view, $\nu$  is characterized as the length of the circle $U(1)\simeq S^1$ (of the unit radius).
\par
Thus it is enough to check  that 
\be 
\label{mt11}
\bf m (\Phi,A)= \frac{\nu}{4\pi}~ \zeta (\Phi,A)
\ee 
for an arbitrary (topologically nontrivial) vacuum YMH configuration $(\Phi^a ({\bf x}), A_\mu^a({\bf x}))$ in the Minkowski space \footnote{Formally, Eq. (\ref{mt11}) permits the half-integer magnetic charge ${\bf m}=1/2$.

Recently the investigation of  BPS monopole solutions involving half-integer magnetic charges ${\bf m}= n/2$ ($n\in \bf Z$) was performed in the paper \cite{Teh}. Unfortunately, in spite of a general attractiveness and interesting  ideas stated in this paper,  such BPS monopole solutions may provoke series of problems at constructing Minkowskian QCD (especially as  the Minkowskian QCD model \cite{David2,David3, Pervush2,LP2,LP1} is in question.

As there was demonstrated in  \cite{Teh},  in the BPS monopole theory involving "half-monopole" solutions, Higgs (vacuum)  BPS monopole modes possesses the spatial asymptotic $\sim (1+r)^{-1}$ at the origin of coordinates. On the other  hand, repeating the arguments  \cite{David3, Pervush2,LP2,LP1}, one can show that Higgs ansatzes $f_{01}^{BPS}(r)$ would, indeed, possess the spatial asymptotic $ f_{01}^{BPS}(0)\to 0$ at the origin of coordinates in order to ensure the ultraviolet asymptotic freedom of quarks  and other effects connected with the discrete vacuum geometry \cite{disc} us adopted for the discussed model.
}

\medskip Let us verify the  equality (\ref{mt11})  in the simple case of the "continuous" vacuum geometry, when $ \Phi ({\bf x})$ takes its values in the vacuum manifold $$R=SU(2)/U(1)=M_0\simeq S^2.$$ 
We think herewith that the surface $\Gamma \subset R$ is topologically equivalent to $S^2$. Note, however, that, in effect, the proof of the formula (\ref{mt1}) does not depend on the choose of the surface $\Gamma \subset R$. On  the other hand, in the case of the "discrete" vacuum geometry \cite{disc}, it is naturally to choose  surfaces $\Gamma$ in such a wise that these are indexed by integers $n\in \bf Z$: more exactly, $\Gamma_n$ can be chosen to coincide with appropriate topological domains in the vacuum manifold.
Then the dependence ${\bf m}(n)$ for magnetic charges  ${\bf m}(n)$, given via  (\ref{potok}),  becomes  greatly transparent also in the case of "discrete" vacuum geometries.

\medskip It is possible to introduce a coordinate system $(\rho,\alpha)$ on $S^2$, with $\alpha$ being the longitude: $0\le \alpha\le 2\pi$, and $\rho$  being the distance  to the south pole: $0\le \rho \le 1$.  For the south pole, $\rho=0$, and for the north one, $\rho=1$; $\alpha$ is arbitrarily in  both the cases \footnote{One may understand the distance from a point on $S^2$ to a pole either as this distance in ${\bf R}^3$ or as the minimal distance on the surface of $S^2$. In both the cases one would normalize the metric in such a wise that the distance between the poles is equal to 1.}. This system of coordinates  generates   coordinates on $\Gamma$. We shall denote them also  as $(\rho,\alpha)$. 

Utilizing the just introduced  coordinates, one may construct the map $f$ of a circle $D^2\in SU(2)\equiv G$ onto $\Gamma$,  mapping the whole its boundary $\partial D^2=S^1$ in a point. Assuming the unit radius for the circle $D^2$, one would then associate the  point of $\Gamma$  with  the coordinates  $(\rho,\alpha)$ to the  point of $D^2$ with  the polar coordinates  $(\rho,\alpha)$ \footnote{Formally, $S^2\cong D^2/\partial D^2=D^2/S^1$. In coordinates,
$$D^2=\{(x,y)\in {\bf R}^2| x^2+y^2\leq 1\};$$
 $$ \partial D^2=S^1=\{(x,y)\in {\bf R}^2| x^2+y^2=1\}.$$ Our task is now to define the map $$f:D^2\to S^2\subset {\bf R}^3 $$ such that 

$\circ $ the internal part of the disc $D^2$ maps bijective onto the sphere $S^2$ without an one point.

$\circ $ the all boundary of the disc $D^2$ maps in this "missing" point.

It is easy to construct such a map. In order that all the boundary $\rho =1$ (in our convention, it is just the north pole!)  collapse into a point, we construct the following map: $$ g(\rho,\phi)=(\sin(\pi \rho) \cos\phi, \sin(\pi \rho) \sin\phi, \cos(\pi \rho)).$$ 
At $\rho=0$, $g(0,\phi)=(0,0,1)$, and this is the north pole. 

At  $0<\rho<1$  we get all points of the sphere except the north and south poles.

At $\rho=1$, $g(1,\varphi) = (0,0,-1)$, and this is just the south pole. The latter result {\it does not depend} on $\varphi$, and this shows that at  $\rho=1$, all the boundary of the disk $D^2$ collapses into the point indeed.

}. \par 

\medskip To prove a one-to-one correspondence between $\pi_2 (R)$ and $\pi_1 (H)$ \footnote{This map guarantees the existence of {\it point topological defects} inside the vacuum manifold $R$. On the other hand, as we shall see below, this is related closely to the proof of Eq. (\ref{mt11}). },  it is necessary (see \S T20 in \cite{Al.S.}) to find the covering map $\beta: D^2 \to G$ for   $\varphi: ~\Gamma  \to R$. \par 
Since $\pi_1 (G)=0$, each map of the circle 
$S^1\simeq U(1)\equiv H$ may be continued into the map $\beta: D^2 \to G$. Considering   a closed way $\tilde \alpha $ in $H$ as a map of $S^1$, one constructs   $\beta: D^2 \to G$ in such a wise that it coincides   with the way $\tilde \alpha $ in $H$ on the boundary of $D^2$ (this boundary is just $S^1$, as we have found out above!).\par 
At a natural map $\pi: G\to G/H=R$ all the residual gauge symmetry group $H$ maps in a point in assuming $H$ being the stationary subgroup for any point of the vacuum manifold $R$. Note that this is a general theoretical-group reasoning! More exactly the latter statement may be written down as
\be \label{jadro}  \pi_1 (H)={\rm ker}~ \pi_1 (G).\ee 
Whence the map $\pi \circ \beta: D^2\to G/H$ maps the whole boundary of $ D^2$ ($\partial D^2=S^1$) into  a  point; therefore it determines \cite{Ryder} an element of the  group $\pi_2 (G/H)$ due to the natural isomorphism  
\be \label{estestv. otobr.}
\pi_2 (G/H)\simeq {\rm ker}~ [\pi_1 (H)\to \pi_1 (G)].
\ee
The isomorphism    (\ref{estestv. otobr.}) has the following origin. Following the monograph \cite{Postn4} (see p. 454 ibid), consider the exact  homotopic sequence  of some fibre bundle with the connected total space $\cal E$ and  base $\cal B$:

 \be
\label{pi1} \pi_0 ({\cal E}) =\pi_0({\cal B}) =0, \ee
having the form
\be
\label{secuenc} ... \to \pi_{n+1}{\cal B}\longrightarrow  \sb{\partial} \pi_n {\cal F} \longrightarrow \sb{ i_* }\pi_n {\cal E}\longrightarrow \sb { p_*} \pi_n {\cal B} \to ...
\ee (where $\cal F$ is the fibre of the considered fibre bundle: $\pi_0({\cal F})=0$). We recommend our readers to study Task 5 p. 455 in \cite{Postn4} in order to prove that the sequence (\ref{secuenc}) is indeed exact. 

\medskip Now, in order to ground the isomorphism    (\ref{estestv. otobr.}), we should consider the principal fibre bundle  
$$ G_{\rm pr}\equiv (G, G/H, H, \pi),$$ where the role of the total space $\cal E$ is played by the initial gauge symmetry group $G$, the degeneration space $R= G/H$ plays the role of the base of that principal fibre bundle, and  the residual gauge symmetry group $H$  plays the role of the typical fibre of that  fibre bundle; $\pi:=G\to R$ is the projection of the principal fibre bundle $ G_{\rm pr}$.

Since, in definition of an exact homotopical sequence \cite{Postn4}, always 
$$ {\rm im}~\alpha ={\rm ker}~ \beta $$ for two neighboring elements $\alpha $ and $\beta $ of this exact homotopical sequence, Eq. (\ref {estestv. otobr.}) is  fulfilled for the principal fibre bundle  $ G_{\rm pr}$.

Further, if $\pi_2 (G)=0$, an element of the  group $\pi_2 (G/H)$ constructed in such a wise does not depend on a choice of the  map $\beta$: it is completely specified by the choice of a homotopical class  of the way $\tilde \alpha $ in $H$. 

Thus we have completely constructed a map 
$\pi_1(H)\to \pi_2 (G/H)$. This map proves to be an isomorphism between  the  groups $\pi_2 (G/H)$ and $\pi_1 (H)$ if and only if

1. the residual gauge symmetry group $H$ is the stationary subgroup for any  point of the vacuum manifold $R=G/H$ \cite{Al.S.}.

2.  In (\ref{secuenc}) we previously set $ \pi_0 ({\cal E}) =\pi_0({\cal B}) =0$. In terms of ${\cal E}=G=SU(2)$, $H={\cal F}=U(1)$ and ${\cal B}=R=SU(2)/U(1)$, this claim is met.  This implies that $\pi_1(H)\to \pi_2 (G/H)$ is isomorphism subtracted from the exact sequence (\ref{secuenc}).

The discrete vacuum geometry worked out in \cite{disc} generates a more complicate situation for the appropriate vacuum manifold. The {\it discrete} factorization \cite{disc,Pervush1}

$$U(1)\cong U_0\otimes {\bf Z}, \quad  \pi_0 (U_0)=0 $$ is applied in order to obtain the nontrivial topological dynamics of the YMH model with vacuum monopole modes on the Gauss law constraint surface.   As a result, one comes to the vacuum manifold
$$ R_{\rm YM}=SU(2)/(U_0\otimes {\bf Z})$$. The analysis carried out in \cite{disc} shows that $\pi_2(R_{\rm YM})={\bf Z}$. Nevertheless, considering  the "long" exact sequence of homotopy groups $\pi_0 \dots \pi_2$ for the initial gauge symmetry group $SU(2)$, the residual gauge symmetry group $U_0\otimes {\bf Z}$  and the vacuum manifold $R_{\rm YM}$ developed in \cite{disc} results the similar isomorphism 

$$ \pi_1( U_0\otimes {\bf Z})=\pi_2(R_{\rm YM})$$ as in the case of the vacuum manifold $R=SU(2)/U(1)$ above! This ensures the existence of point (hedgehog) topological defects in both the cases: in the "continuous" as well as in the "discrete" one.
%Formally, $ \pi_0 [ U_0\otimes {\bf Z}]={\bf Z}$. 

\medskip In  terms of the  isomorphism (\ref{point1}) (us analyzed above), i.e. $\pi_2 (G/H)=\pi_1 (H)$, the problem to construct the covering map $\beta: D^2 \to G$ for a 
$\varphi: ~\Gamma  \to R$ comes  to the relation \be  \label{varphi} \varphi (\rho,\alpha)=T (\beta^{-1} (\rho,\alpha)) \varphi_0, 
\ee
where $\rho$ and $\alpha$ on the left-hand side of this equality are above introduced coordinates on $\Gamma$,
while on the right-hand side stand polar coordinates on $D^2$; the symbol $\varphi_0$ means the value of the field  $\varphi$ in the north pole of the surface $\Gamma$ (i.e. in the point where $\rho=1$ and $\alpha$ is arbitrarily); $T$ stands for a representation  of the initial gauge symmetry group $G$ \footnote{If we once again consider the principal fibre bundle  $G_{\rm pr}$, its projection $\pi:=G\to R$ associates the point
$$ \pi (g)= T(g^{-1})\varphi_0\in R=G/H$$  to the point $g\in G$.}.

As the consequence of (\ref {varphi}), 
$$ T (\beta^{-1} (1,\alpha)) \varphi_0 =\varphi_0.$$ This means that the map $\beta^{-1}$ can be treated as a map $S^1\to H$ on the boundary  of   $D^2$, where now $H$ is the stationary subgroup of the element $\varphi_0$ (according to our convention).\par
Associating the element of the group $\pi_1(H)$ specified by this relation to the appropriate element of the group $\pi_2(R)$ determined by the map $\varphi: ~\Gamma  \to R$, we again come to the   one-to-one correspondence between $\pi_1(H)$ and $\pi_2(R)$. 

\medskip  Thus in the case $H=U(1)$, that is our present interest,  the topological  number of the Higgs field $\Phi$ may be calculated by the formula 
\be  \label{ksig}
 \zeta (\Phi,A) =\frac {1}{\nu}\int \limits _0^{2\pi}\frac {\partial \lambda}{\partial\alpha} d \alpha = \frac {1}{\nu} (\lambda (2\pi)- \lambda (0)), 
\ee with  $\lambda (a)$ being the continuous function setting by the relation \footnote{The $-$ sign for $\beta(1,\alpha)$ below is the question of convention only. The choice $-$ in \cite{Al.S.} corresponds to the right action of $U(1)$.} $$ \beta(1,\alpha)=\exp (-\lambda(\alpha)h) $$ (each element of  the circle $H=U(1)$ may be written down as $\exp (\pm \lambda h)$, where $0\leq \lambda \leq \nu$. Considering then
$2\pi \nu^{-1}\lambda$ as the angular coordinate on $H$ \footnote{Indeed, the circle $U(1)\cong S^1$ is parametrized through the angular coordinate $\theta \in [0,2\pi]$, while in our consideration $\lambda \in [0,\nu]$. To go over from $\lambda$ to $\theta$, we write
$$ \theta =\frac {2\pi}\nu  \lambda  \in [0,2\pi].$$
}, we come to (\ref{ksig}). $\zeta (\Phi,A) \in {\bf Z}$. To demonstrate this fact, we assume that $\beta(1,\alpha)$ is a loop in $U(1)$. Then 
$$ \beta(1,0)=\beta(1,2\pi),  $$ or $$\exp(-\lambda(2\pi)h) =\exp(-\lambda(0)h).$$ Hence  $$\zeta (\Phi,A)= (\lambda(2\pi)-\lambda(0))/\nu =k \nu/\nu\in {\bf Z}. $$

\medskip To get now Eq.  (\ref{mt1}), performing a gauge transformation  with the function $\beta(\rho,\alpha)$, one can transfer the Higgs field $\Phi$ into the field equal to $\varphi_0$ in the whole surface $\Gamma$, with the exception of the north pole (now $(\rho,\alpha)$ are treated as coordinates on $\Gamma$; as $\rho<1$, these coordinates determine a one-to-one correspondence between the interior of the circle $D^2$ and points of $\Gamma$  with the removed north pole). 
\par 
This means that, upon such gauge transformations, the Higgs field becomes     
\be 
\label{sfi}  
\sigma \Phi ({\bf x})= \varphi_0, \ee 
with $\sigma $ being the map associated to an arbitrary value of the Higgs field $\Phi ({\bf x})$ the nearest to it point of the vacuum manifold $R$. In turn, the given YM field $A_\mu$  becomes $A'_\mu$   upon such gauge transformations. The field $A'_\mu$ takes its values in the Lee algebra of $H\subset G$, i.e. $A'_\mu=a_\mu h$. The field $a_\mu$, in turn, may be treated as an "electromagnetic"  field; then we may come to Eqs. (\ref {H}) for the YM "magnetic"  tension and  (\ref{mul}) for the magnetic charge. 

If the investigated fields are regular in  the north pole, one may put out a small neighborhood $\epsilon$ of the north pole in such a wise that the "magnetic"  flux (\ref{mul}) remains almost unaltered.  More precisely, if we denote as $\Gamma_\epsilon $ the part of the surface $\Gamma$  for which  $\rho\leq 1-\epsilon$, then the flux through  $\Gamma_\epsilon $ approaches   the flux through  $\Gamma$ as $\epsilon \to 0$.                 The boundary of $\Gamma_\epsilon $ is a small circle $L_\epsilon $. 
Thus  \bea \label{mul1} {\bf m}&=& \frac{1}{4\pi}\lim \limits _{\epsilon \to 0}\int d{\bf S}~ {\bf H}= \nonumber 
\\ \frac{1}{4\pi}\lim \limits _{\epsilon \to 0} \oint a_\mu d x ^\mu &=& 
\frac{1}{4\pi}\lim \limits _{\epsilon \to 0} \int \limits _0^{2\pi}
a_\alpha (1-\epsilon, \alpha) d \alpha.  \eea
The symbols $a_\rho (\rho, \alpha)$, $a_\alpha (\rho, \alpha)$ stand here for the components of the vector potential $ a_\mu$ in the coordinates $(\rho, \alpha)$ on $\Gamma$. 

Let us suppose for simplicity that the YM field $ A_\mu$ becomes zero closely about the north pole. Then we obtain the  field 
$$ A'_\alpha (\rho, \alpha) = a_\alpha (\rho, \alpha) h = -\frac{\partial}{\partial_\alpha}[\beta ^{-1}(\rho, \alpha)]~ \beta (\rho, \alpha), \quad \rho \geq 1-\epsilon. $$

As  $$\beta (\rho, \alpha)\to \beta (1, \alpha)=\exp (-\lambda(\alpha)h)$$
at $\rho \to 1$, 
then (since we deal with the Abelian group $U(1)$ now)
$$ a_\alpha (\rho, \alpha) \to \partial \lambda / \partial \alpha, $$  
and, because of (\ref{mul1}), 
\be  
\label{mt2} 
{\bf m}=\frac{1}{4\pi} (\lambda (2\pi)- \lambda (0)). 
\ee  Comparing (\ref{mt2}) and (\ref {ksig}), we come to (\ref{mt1}).\par

\bigskip Let us now show that, in the considered Minkowskian YMH theory involving vacuum BPS monopole modes (as well as in other YM theories with monopoles \cite{Wu,H-mon, Polyakov, Hooft}) electric and magnetic charges can be quantized by Dirac \rm \cite{Ryder,Cheng}.

Really, 
 electric charges of particles corresponding to a field $\Phi$ are determined by the formula $-i\lambda_k$, with $\lambda_k$ being eigenvalues of the operator $t(h)$ ($t(h)$ is the given representation of the Lee algebra of the group $H$).  The operator $t(h)$ is anti-Hermitian; thus its eigenvalues are imaginary.  More precisely, if $\Phi=\sum \limits_k  \Phi^k f_k$, where $f_1,\dots f_n$ are eigenvectors of the operator $t(h)$ and if only the "electromagnetic" part of the gauge field $A_\mu$: $a_\mu$, is different from zero (i.e. $ A_\mu=a_\mu h$), then 
$$\nabla _\mu \Phi =\partial_\mu \Phi + t(a_\mu h) \Phi =\sum \limits_k (\partial_\mu \Phi^k + \lambda_k a_\mu \Phi^k )f_k.$$ 
It  follows from the relation $\exp (\nu h)=1$ that 
 $$ T(\exp (\nu h))= \exp (\nu~ T(h))=1.$$
Therefore for all the eigenvalues $\lambda_k$ of the operator $t(h)$ the equality $\exp (\nu h)=1$ is satisfied. Whence $$ \nu \lambda_k = 2\pi n i, \quad n\in {\bf Z};$$ i.e. electric charges are equal to $2\pi n/\nu$. 
Thus these charges are integer multiples of the number $2\pi \nu^{-1}$. 

Since magnetic charges are integer multiples of the number $\nu (4\pi)^{-1}$,  the product of the electric charge $e$ of a particle with the magnetic charge $\bf m$ of (another) particle is a 
half-integer: 
 \be 
\label{dirk}
 e{\bf m}= \frac{1}{2} n, \quad n \in {\bf Z}. 
\ee  In the calculations \cite{Al.S.} about the magnetic charge just performed we have used the normalization of the YMH Lagrangian density in (\ref {YM L}) involving the coefficient $-1/(4g^2)$ in front of  $F_{\mu \nu}^2$. 
But this normalization is not standard. Usually, one utilizes the standard normalization with the coefficient -1/4 (instead of $-1/(4g^2)$) standing therein. In this case electric charges $e$ can be chosen to be  
\be \label{eg} e=-i\lambda_k g=\frac{2\pi n}{\nu} g ,\ee 
and magnetic  charges ${\bf m}$, 
\be \label{mg} {\bf m}=\frac{\nu}{4\pi g}~\zeta; \quad \zeta \in {\bf Z}.\ee  
Then the standard relation (\ref {dirk}) remains immovable.

Thus there exists a one-to-one correspondence between 
the magnetic  and  topological charges. 
As a consequence of Eq. (\ref {eg}) \cite{Al.S.}, the YM coupling constant $g$ can be expressed through an electric charge $e$:
\be \label{eg1}
g=\frac{\nu e}{2\pi n}, \quad n\in {\bf Z}. 
\ee
Alternatively, one can  express $g$ through an magnetic  charges ${\bf m}$:
\be \label{mg1} 
1/g=\frac{4\pi{\bf m}}{\nu \zeta}, \quad\zeta\in {\bf Z}. 
\ee
Latter Eq. shows that even if $\zeta=0$, $1/g$ is different from zero and can take finite values.
    
\bigskip
The analysis has been performed in the papers \cite{David2,David3, Pervush2,LP2,LP1,fund}, devoted to the Dirac fundamental quantization  \cite{Dir} of  Minkowskian (Higgs) models, shows that just nonzero vacuum "electric" fields $F_{0i}^a$ in the Minkowskian YMH model \cite{David2,David3, Pervush2,LP2,LP1} quantized by Dirac are associated with a nontrivial (topological) dynamics inherent in that model.

Herewith mentioned nonzero vacuum "electric" fields $F_{0i}^a$ in the Minkowskian YMH model \cite{David2,David3, Pervush2,LP2,LP1} quantized by Dirac take the shape of vacuum "electric" monopoles, directly proportional to the topological  dynamical variable $\dot N(t)$, the time derivative of $ N(t)$, the non integer degree of the map
referring to the $U(1)\to SU(2)$ embedding:
\be
 \label{el.m.}
  F^a_{i0}={\dot N}(t)D ^{ac}_i(\Phi_k ^{(0)}) \Phi_{(0)c}({\bf x}).        \ee
Topologically trivial vacuum "electric" monopoles $ F^a_{i0}$ implicate Higgs vacuum BPS \linebreak monopole solutions $\Phi_{(0)a}({\bf x})$.

As it was explained in Ref. \cite {Pervush1},  this is the immediate result of choosing temporal components of YM vacuum modes in the Minkowskian YMH model \cite{David2,David3, Pervush2,LP2,LP1} quantized by Dirac to be proportional to Higgs vacuum BPS monopoles $\Phi_{(0)a}({\bf x})$: with $\dot N(t)$ playing the role of the proportionality coefficient:
\be \label{Za} Z^a=\dot N(t) \Phi^{(0)a}({\bf x}).\ee
This choice of  YM vacuum modes is dictated just by the specific of the Dirac fundamental scheme \cite {Dir}, in which $ Z^a $ are (general) solutions to the YM Gauss law constraint
\be
 \label{odnorod}
 (D^2)^{ab} \Phi_{b(0)}=0.
\ee
This, in turn, is the look of the YM Gauss law constraint (\ref{Gau}):
\be
\label{Gauss YM}
\partial W/\partial A_0=0 \Longleftrightarrow [D^2(A)]^{ac}A_{0c}= D^{ac}_i(A)\partial_0 A_{c}^i,
\ee
resolved with the Coulomb covariant gauge \cite{David2,Pervush2,fund}
\be
\label{kg}
D^{ac}_i(A)\partial_0 A_{c}^i=0
\ee
for YM fields $A$.

As it was explained in the papers \cite{David2,David3, Pervush2,LP2,LP1, fund}, the Coulomb covariant gauge (\ref{kg}) may be satisfied by {\it topological Dirac variables} \cite{David2,David3, Pervush2, LP1, fund}
\be \label{per.Diraca} {\hat A}^D_i({\bf x},t):=v^{T(n)}({\bf x},t)({\hat A}_i+\partial_i)(v^{T(n)})^{-1}({\bf x},t); \quad {\hat A}_i= g \frac {\tau ^a}{2i}A_{ai};\quad  n\in {\bf Z};
\ee
implicating Gribov topological multipliers $ v^{T(n)}({\bf x},t)$ (in the Minkowskian Higgs model with vacuum BPS monopole solutions, these topological multipliers depend explicitly on a scalar combination of Higgs BPS monopoles via $\tau^a \Phi_a$; $\tau^a$ ($a=1,2,3$) are Pauli matrices). \par
The functionals ${\hat A}^D_i({\bf x},t)$ specified in such a wise prove to be gauge invariant and transverse fields:
\be \label{Dv} \partial_0 D_i {\hat A}^D_i({\bf x},t) =0; \quad  u ({\bf x},t)  {\hat A}^D_i({\bf x},t)   u ({\bf x},t)^{-1}= {\hat A}^D_i ({\bf x},t)    \ee
for gauge matrices $ u ({\bf x},t)$. \par
\medskip
As it was discussed in the review \cite{fund} (repeating the said in Refs. \cite{David2, Pervush2,LP2,LP1}), vacuum "electric" monopoles (\ref{el.m.}) involve  collective "solid" rotations of the Minkowskian physical YMH BPS monopole vacuum. \par
These may be described by the action functional \cite{David2}
\be
\label{rot}
W_{\rm coll}=\int d^4x \frac {1}{2}(F_{0i}^c)^2 =\int dt\frac {{\dot N}^2(t) I}{2},
\ee
with
\be
\label{I1}
I=\int \sb {V} d^3x (D_i^{ac}(\Phi_k)\Phi_{(0)c})^2 = \frac {4\pi^2\epsilon}{\alpha _s} =\frac {4\pi^2}{\alpha _s}\frac {1}{ V<B^2>} \ee
being the {\it rotary momentum} of the Minkowskian YMH vacuum. \par
In Eq. (\ref{I1}), $\Phi_k $ are vacuum BPS monopole modes; $\alpha_s\equiv g^2/4\pi$ (with $g$ being the YM coupling constant); $V$ is the spatial volume. \par
The purely real spectrum
\be \label{Pcr}
P_N\equiv \frac{\partial W_{\rm coop}}{\partial\dot N}={\dot N} I= 2\pi k +\theta,
\ee
with the $\theta$-angle chosen to vary in the interval $[-\pi, \pi]$ \cite {David2}.

\medskip
At the FP "heuristic" quantization \cite {FP1} of Minkowskian Higgs models with monopoles, all these and similar vacuum rotary effects disappear with disappearing "electric" fields $ F^a_{i0}$. \par
\medskip
The nontrivial topological dynamics inherent in the Minkowskian Higgs model \cite{David2,David3, Pervush2,LP2,LP1} (involving vacuum  BPS monopole solutions) quantized by Dirac \cite {Dir} draws a peculiar watershed between the FP "heuristic" \cite {FP1} and "fundamental" \cite {Dir} approaches to the quantization of gauge models. \par
This is associated \cite {fund,Arsen} with violating the gauge equivalence (independence) theorem \cite{Taylor, Slavnov1} in the case of collective vacuum excitations. \par
In the Minkowskian YMH model \cite{David2,David3, Pervush2,LP2,LP1} quantized by Dirac \cite {Dir}, there exist such "collective vacuum excitations". \par
There are just zero mode solutions $Z^a$, (\ref{Za}), to the YM Gauss law constraint (\ref {odnorod}), (\ref{Gauss YM}), generating various vacuum rotary effects in the enumerated model.
\section{ Superfluid properties of  Minkowskian Higgs model involving vacuum  BPS monopole modes and "continuous" vacuum geometry.  }
The "classical" Minkowskian Higgs model \cite{Al.S.,BPS,Gold} with vacuum  BPS monopole solutions proves to be the unique model with monopoles in which the appropriate vacuum possesses the manifest superfluid properties.

Now we shall attempt to argue in favour of  this statement.

\medskip
In the recent paper \cite{LP1} there was pointed out the role of the Bogomol'nyi  equation (\ref{Bog}) as the {\it potentiality condition} for the Minkowskian YMH vacuum  involving BPS monopole modes.

Let us interpret the latter assertion.

Mathematically, any potentiality
condition may be written down as
\be \label{potcon} {\rm rot}  ~{\rm grad} ~{ \Phi}=0 \ee
for a scalar field $\Phi$ (to within a constant).\par
Thus any  potential field may be represented as ${\rm grad} ~{ \Phi}$ (to within a constant). \par
It is a good prompt for us. \par
In the Minkowskian YMH theory involving BPS monopole solutions (for instance, in the "classical" Minkowskian Higgs model \cite{Al.S.,BPS,Gold} with  BPS monopoles), there exists always such a scalar field. It is just the Higgs (world) scalar $\Phi$ represented as the Higgs BPS monopole in the vacuum sector of that theory. \par
Then it is easy to guess  that the Bogomol'nyi equation (\ref{Bog}), having the look (\ref{potcon}), can be treated  as  the  potentiality condition for the Minkowskian YMH vacuum involving vacuum BPS  monopole solutions. It is so due to the Bianchi identity $D B=0$ which can be represented as $$\epsilon ^{ijk}\nabla _i F_{jk}^b =0$$ (at neglecting the items in $DB$ directly proportional to $g$ and $g^2$).\par
\medskip
Indeed, there can be drawn a highly transparent parallel between the  Minkowskian YMH vacuum involving vacuum BPS  monopole solutions (say, \cite{Al.S.,BPS,Gold}) and a liquid helium II specimen described in the Bogolubov-Landau model \cite {N.N.}. \par 
In the latter case, the potential motion is proper to the superfluid component in  this liquid helium  specimen. \par
The superfluid motion in a liquid helium II is the motion without  a friction between the superfluid component and the walls of the vessel where a liquid helium  specimen is contained. \par
Thus the viscosity of the superfluid
component in a helium II  is equal to zero, and vortices (involving ${\rm rot}~ {\bf v}\neq 0$) are absent in the superfluid component of a helium. \par
As L. D. Landau   showed \cite {N.N.}, at velocities of the liquid exceeding a {\it critical velocity} $ v_0= {\rm min}~ (\epsilon/p)$ for the ratio of the  energy $\epsilon$ and momentum $p$ for quantum excitations spectrum in
the liquid helium II, the dissipation of the liquid helium energy occurs via arising excitation quanta with momenta $\bf p $ directed antiparallel to the velocity vector $\bf v$. Such dissipation of the liquid helium energy becomes advantageous \cite{Halatnikov}  just at
$$ \epsilon+ {\bf p~v}<0 \Longrightarrow \epsilon -p~v<0.$$
From the above reasoning concerning properties of potential motions, it becomes obvious that the vector ${\bf v}_0$ of the critical velocity for the superfluid potential motion possesses the zero curl: ${\rm rot}~ {\bf v}_0=0$. \par
In this case,  according to  (\ref {potcon}), the critical velocity ${\bf v}_0$ of the superfluid potential motion in a liquid helium specimen may be represented \cite{Landau52} as 
\be \label{alternativ} {\bf  v}_0 =\frac{\hbar}{m} \nabla \Phi(t,{\bf r}),
\ee
where $m$ is the mass of a helium atom and $\Phi(t,{\bf r})$
is the phase of the complex-value helium Bose condensate wave function $\Xi (t,{\bf r})\in C$.
\par
Note that the latter one may serve as a complex order parameter in the Bogolubov-Landau  model of the liquid helium \cite{N.N.}; its explicit look is \cite{Landau52}
\be
\label{Xi1}
 \Xi (t,{\bf r})= \sqrt {n_0(t,{\bf r})}~ e^{i\Phi(t,{\bf r})},
\ee
with $ n_0(t,{\bf r})$ being the number of particles in the ground energy state $\epsilon=0$.\par
Thus the similar look for the vacuum "magnetic" field $\bf B$ in the Minkowskian Higgs model involving   BPS monopole solutions, generating by the Bogomol'nyi equation (\ref{Bog}), and for the critical velocity ${\bf v}_0$ of the superfluid motion in a liquid helium II, given by Eq. (\ref{alternativ}), testifies in favour of the  potential motions occurring therein. \par
In this case, drawing a highly transparent parallel between the  Minkowskian YMH vacuum  involving BPS monopole solutions \cite {LP2,LP1,Al.S.,BPS, Gold}  and a liquid helium II specimen described in the Bogolubov-Landau model \cite {N.N.}, we can also conclude about manifest superfluid properties of the Minkowskian YMH vacuum  involving BPS monopoles. \par
As in the Bogolubov-Landau model \cite {N.N.} of  liquid helium II, the ground cause of the superfluid properties of the
Minkowskian YMH vacuum with BPS monopoles roots in  long-range correlations of local excitations \cite {Pervush1}.\par 
While in the Bogolubov-Landau model \cite {N.N.} of  liquid helium II this comes to repulsion forces between helium atoms as the cause of superfluidity effects, in the  Minkowskian YMH vacuum involving BPS monopole solutions, the cause of the superfluidity taking place is in the strong YMH coupling $g$ (entering effectively the appropriate action functional (\ref {YM L})). \par
\medskip
The chief thing in alike superfluid effects occurring in a liquid helium II specimen as well as in the Minkowskian YMH vacuum involving BPS monopoles is that these both physical systems are {\it non-ideal gases}. \par
In ideal gases no superfluidity phenomena are possible. \par
There can be demonstrated (see e.g. \S4 of Part 6 in  \cite{Levich})  that in ideal gases a deal of particles is accumulated on the  zero energy quantum level at temperatures $T<T_0$; herewith the  temperature $T_0$
\be
\label{condensation temperature}
kT_0= \frac{1}{(2.61)^{2/3} }\frac{h^2}{2\pi m} (\frac{N}{V})^{2/3}
\ee
(with $k$ and $h$ being, respectively, the Boltzmann and Planck constants; $N$ being the complete number of particles; $V$  being the volume occupied by the
ideal Bose gas; $m$  being the mass of a particle)
 is called \it the condensation temperature\rm, while the above  deal of particles is called \it the Bose  condensate\rm. \par
\bigskip
Now we should like argue in favour that only the Minkowskian Higgs model with vacuum BPS monopole solutions possesses the above described superfluid properties, distinguishing it from another Minkowskian Higgs models with vacuum BPS monopoles. \par
In our argumentation we shall follow the paper \cite {Hooft}.\par
Consider again the 't Hooft-Polyakov model \cite {H-mon, Polyakov}. \par
As it is well known \cite {Ryder}, $D_i \phi^a\to 0$ as $ r\to \infty$ for a
't Hooft-Polyakov Higgs monopole $\phi^a $ (having the look (\ref{hedg})).\par
In this case, asymptotically,
\be \label{3.5}
{\bf B}^a_i D^i\phi_a ={\partial}_i({\bf B}^i_a\phi_a)=0,
\ee
because of the Bianchi identity
$$ D_i{\bf B}_i^a\equiv\frac{1}{2} \epsilon_ {ijk} D_i F_{jk}^a=0
$$
and the remark \cite {Al.S.} that ${\bf B}^i_a\phi_a$ is a
$U(1)\subset SU(2)$ scalar;  thus one can replace the covariant derivative $D$ with the partial one, $\partial$, for ${\bf B}^i_a\phi_a$. \par
In turn, the complete energy of the YMH configuration may be represented as \cite{Al.S., Hooft}
\be \label{complet}
E_{\rm compl} = \int d^3x~ [\frac{1}{2}(D\phi_a \pm {\bf B}_a)^2+ \frac{\lambda}{4}((\phi^a)^2- a^2)]+ \frac{4\pi}{g^2} M_W.
\ee
The last item in Eq. (\ref{complet}) involves the mass $ M_W $ of the
$W$-boson. \par
Such look of $ E_{\rm compl} $ originates from the paper \cite {H-mon} devoted to the 't Hooft-Polyakov model. \par
The connection between the energy integral $ E_{\rm compl} $ and the general action functional (\ref {YM L}) \cite {LP2,LP1} of the Minkowskian Higgs model is given by the identity \cite{Hooft}
\be \label{togd}
(D\phi_a)^2+{\bf  B}_a^2=
(D\phi_a \pm {\bf B}_a)^2 \mp 2 {\bf B}_a D\phi_a.
\ee
Herewith the last item on the right-hand side of (\ref{togd}) vanishes at the spatial infinity, as we have noted above \cite {Ryder}.\par
Just from Eq. (\ref{complet})  one can read formally the Bogomol'nyi equation in the shape (\ref{Bog}).
\par
In the 't Hooft-Polyakov model \cite {H-mon, Polyakov} the Bogomol'nyi equation (\ref{Bog}) determines the Bogomol'nyi bound \cite{Hooft}
\be \label{Emin1}
M_{\rm mon} =\frac{4\pi}{g^2} M_W
\ee
for the complete energy $ E_{\rm compl}$, (\ref{complet}), of the YMH configuration at going over to the BPS limit (\ref{lim}) \cite{BPS}.
\par
Then  the asymptotic  $D_i\phi^a\to 0$ as $ r\to \infty $ \cite{Ryder} for 't Hooft-Polyakov monopoles \cite {H-mon, Polyakov} forces to vanish identically the first item under the integral sign in $ E_{\rm compl}$ ($\vert {\bf B}\vert =0$).\par
In the light of the said above it becomes obvious that the vacuum "magnetic" field $\bf B$, playing the role of the (critical) velocity for the superfluid motion in the Minkowskian non-Abelian vacuum with BPS monopoles, actually approaches zero in the 't Hooft-Polyakov model \cite {H-mon, Polyakov}, involving the $D_i\phi^a\to 0$ as $ r\to \infty $ asymptotic \cite{Ryder} for Higgs monopoles.  \par
\medskip
There is no superfluidity also in the Wu-Yang model \cite {Wu}.\par
The absence of Higgs vacuum modes in that "purely YM" Minkowskian model is the cause of such situation. \par
On the other hand, Wu-Yang monopoles \cite {Wu} approximate good, at the spatial infinity, YM BPS monopole solutions \cite {LP2,LP1, Al.S.,BPS, Gold}, ensuring manifest superfluid properties of the appropriate YMH vacuum \footnote{Indeed, Wu-Yang monopoles \cite {Wu} approximate YM ones outside cores of latter monopoles \cite {Pervush2,LP2}.}.\par
\section{Discussion.}
In the present study we have discussed the specific of  various Minkowskian (Higgs) models involving vacuum  monopole solutions. \par
Our initial premise was herewith assuming about the "continuous", $\sim S^2$, (\ref{contin}), vacuum geometry in these models. \par
There was demonstrated that this assuming confines itself with  the FP "heuristic" quantization \cite{FP1} of Minkowskian (Higgs) models involving vacuum  monopoles. \par
Moreover, the absence of "electric" fields $F_{0i}^a$ in these models prevents any (topologically nontrivial) dynamics via fixing the Weyl (temporal) gauge $A_0=0$ through the $\delta (A_0)$ multipliers in appropriate FP integrals. \par
The opposite situation can be observed at the Dirac fundamental quantization \cite{Dir} of Minkowskian Higgs models, coming to the Gauss-shell reduction of these models with choosing a definite (say, rest \cite {David2,Pervush2}) reference frame. \par
As it was discussed in Refs. \cite {David2,Pervush2, David3,LP2,LP1,fund} with the example of the Minkowskian Higgs model involving vacuum BPS monopole solutions, solving the YM Gauss law constraint (\ref{Gau}), (\ref {Gauss YM}) with the transverse Coulomb covariant gauge (\ref {kg}) implies temporal components $Z^a$ \cite {Pervush1}, (\ref {Za}), of YM fields (referring to the BPS monopole vacuum). \par
These, in turn, generate vacuum "electric" monopoles (\ref {el.m.}) and the action functional (\ref {rot}), describing correctly collective solid rotations of the Minkowskian YMH BPS monopole vacuum. \par
\medskip
Utilizing the general theory of topological defects (stated good in the monograph \cite {Al.S.}), we have detected point hedgehog  topological defects in all the Minkowskian Higgs models involving vacuum  monopoles and the continuous geometry (\ref{contin}) of appropriate vacuum manifolds. \par
There was also shown that the Minkowskian Higgs model involving BPS monopole solutions is the unique model in which superfluid vacuum phenomena are possible. \par
The manifest superfluid properties of that model are determined by the Bogomol'nyi  equation (\ref {Bog}), derived \cite {Al.S.} at evaluating the (topologically degenerated) Bogomol'nyi bound $E_{\rm min}$, (\ref {Emin}), of  the YMH energy in the Minkowskian Higgs model with BPS monopoles. \par
The essential point of that deriving was going over to the BPS limit (\ref{lim}) {\cite {Al.S.,BPS}. \par
The Bogomol'nyi  equation (\ref {Bog}) may be interpreted as the potentiality condition for the BPS monopole vacuum. \par
Herewith  the transparent parallel between this vacuum and the superfluid component in a liquid helium II specimen \cite {N.N.} is on hand. \par
\bigskip
On the other hand, the Bogomol'nyi  equation (\ref {Bog}) and manifest superfluid properties of the Minkowskian Higgs model with BPS monopoles (generated by this equation)  prove to be compatible with the Dirac fundamental quantization \cite {Dir} as well as with the FP "heuristic" one \cite {FP1} of that model. \par
In the next paper of the series we devote to remarks about the "Minkowskian Higgs model quantized by Dirac" we shall attempt to argue in favour of the latter statement. \par
Our principal idea will be here that manifest superfluid properties of the Minkowskian Higgs model with BPS monopoles quantized by Dirac are determined by the "potentiality condition" \cite {Pervush1}
\be \label{p. con}  D^2\Phi=0     \ee
imposed onto the Higgs field $\Phi$ having the look of a vacuum BPS monopole. \par
But this "potentiality condition" comes to the Bogomol'nyi  equation (\ref {Bog}) due to the Bianchi identity
$$ D B=0.   $$
It will be shown, repeating the arguments \cite {Pervush2,LP2,LP1,Arsen}, that just Eq. (\ref {p. con}) specifies the ambiguity  in the choice of the Coulomb covariant gauge (\ref {kg}) for (topologically trivial) YM fields (\ref {per.Diraca}), that are, indeed, topological Dirac variables: gauge invariant and transverse.

In  \cite {Pervush2,LP2,LP1}, Eq. (\ref {p. con}) was referred to as the {\it Gribov ambiguity equation}.

In this way, the connection between the Gauss-shall reduction of the Minkowskian Higgs model with BPS monopoles and manifest superfluid properties of that model will be ascertained.

\medskip
The next important topic of the coming investigations will be specifying the explicit look of stationary Gribov topological multipliers
$$v^{T(n)}({\bf x})= v^{T(n)}({\bf x},t) \vert_{t=t_0},$$
entering topological Dirac variables (\ref {per.Diraca}) in the fixed (initial) time instant $ t=t_0$.

Herewith we shall concentrate our efforts about the behaviour of Gribov topological multipliers $ v^{T(n)}({\bf x})$ at the spatial infinity ($\vert {\bf x}\vert \to \infty$).

The principal result will be demonstrated is \cite {Arsen, Pervush1}
$$  v^{T(n)}({\bf x}) \to 1\quad {\rm as}~  \vert {\bf x}\vert \to \infty.    $$
We shall follow the paper \cite {Azimov} at grounding this fact.

As it was noted in Refs. \cite {fund, Azimov}, such asymptotical  the behaviour of Gribov topological multipliers $ v^{T(n)}({\bf x})$ at the spatial infinity provides the infrared (topological) confinement of these multipliers in gluonic and fermionic (quark) Green functions in all the orders of the perturbation theory.

\end{document}